\documentclass[12pt,preprint]{aastex}
\usepackage{epsfig}

\begin{document}

\def\gax    {${_>\atop^{\sim}}$}
\def\lax    {${_<\atop^{\sim}}$}
\def\chandra {{\it Chandra}~}
\def\xmm     {{\it XMM-Newton}~}
\def\etal   {{\it et~al.}~}
\def\msun   {{$M_{\odot}$}}
\newcommand{\kms}{\ifmmode~{\rm km~s}^{-1}\else ~km~s$^{-1}~$\fi}
\newcommand{\atoms}{\ifmmode{\rm ~atoms~cm^{-2}} \else ~atoms cm$^{-2}$\fi}
\newcommand{\aox}{\ifmmode{\alpha_{\rm ox}} \else $\alpha_{\rm ox}$\fi} 
\newcommand{\cmsq}{\ifmmode{\rm ~cm^{-2}} \else cm$^{-2}$\fi}
\newcommand{\nhgal}{\ifmmode{ N_{H}^{Gal}} \else N$_{H}^{Gal}$\fi}
\newcommand{\nhintr}{\ifmmode{ N_{H}^{intr}} \else N$_{H}^{intr}$\fi}
\newcommand{\nhtot}{\ifmmode{ N_{H}^{tot}} \else N$_{H}^{tot}$\fi}
\newcommand{\dM}{$\dot M$}
\newcommand{\dMM}{$\dot{M}/M$}
\newcommand{\dMedd}{$\dot M_{\rm Edd}$}
\newcommand{\pl}{$\pm$}
\newcommand{\nh}{$N_{\rm H}$}
\newcommand{\sig}{$\sigma$~}
\newcommand{\mbh}{$M_{\rm BH}$}

\title{Supermassive black holes, pseudobulges, and the narrow-line
Seyfert 1 galaxies}

\author{Smita Mathur\altaffilmark{2}}
\affil{Astronomy Department, The Ohio State University, Columbus, OH 43210} 
\email{smita@astronomy.ohio-state.edu}

\author{Dale Fields}
\affil{L.A. Pierce College, Woodland Hills, CA, 91371}
\email{FieldsDL@piercecollege.edu}

\author{Bradley M. Peterson\altaffilmark{2}}
\affil{Astronomy Department, The Ohio State University, Columbus, OH 43210} 
\email{peterson@astronomy.ohio-state.edu}


\author{Dirk Grupe} 
\affil{Pennsylvania State University, State College, PA} 
\email{grupe@astro.psu.edu}

\altaffiltext{2}{Center for Cosmology and Astro-Particle Physics, The
 Ohio State University, Columbus, OH 43210}

\begin{abstract}
We present HST/ACS observations of ten galaxies that host narrow-line
Seyfert 1 (NLS1) nuclei, believed to contain relatively smaller mass
black holes accreting at high Eddington ratios.  We deconvolve each ACS
image into a nuclear point source (AGN), a bulge, and a disk, and fitted
the bulge with a Sersic profile and the disk with an exponential
profile. We find that at least five galaxies can be classified as having
pseudobulges. All ten galaxies lie below the \mbh--L$_{bulge}$ relation,
confirming earlier results. Their locus is similar to that occupied by
pseudobulges. This leads us to conclude that the growth of BHs in NLS1s
is governed by secular processes rather than merger-driven. Active
galaxies in pseudobulges point to an alternative track of black
hole--galaxy co-evolution. Because of the intrinsic scatter in black
hole mass--bulge properties scaling relations caused by a combination of
factors such as the galaxy morphology, orientation, and redshift
evolution, application of scaling relations to determine BH masses may
not be as straightforward as has been hoped.
\end{abstract}

\keywords{galaxies:active---galaxies:nuclei---galaxies:spiral}

\section{Introduction}

Active galaxies are ``active'' because they accrete matter onto their
supermassive black holes. However, whether this accretion leads to a
significant growth of the nuclear black hole has been a matter of some
debate. Results on the X-ray background and the better determination of
the local black hole mass density have led to the conclusion that
indeed, most of the black hole growth happens during the active phase
(e.g. Barger et al. 2001, Aller \& Richstone 2002, Yu \& Tremaine 2002,
Graham \& Driver 2007).

The mass of the nuclear black hole, \mbh, and the bulge luminosity of
its host galaxy appear to be correlated (Magorrian et al. 1998).  The
correlation between \mbh\ and the bulge velocity dispersion \sig\ is
also observed to be very tight with scatter of only 0.3dex (Gebhardt et
al. 2000a, Ferrarese \& Merritt 2000, Merritt \& Ferrarese 2001) with
log \mbh~= $a+b\times$log ($\sigma_*/\sigma_0)$ with \mbh~ in units of
\msun~ and $\sigma_0$=200 \kms, $b=4.02$ and $a=8.13$ (Tremaine et al.
2003). Basically, the mass of the black hole seems to be correlated with
the mass of the bulge (H\"{a}ring \& Rix 2004). Interestingly, the above
relation for normal galaxies also extends to active galaxies (Gebhardt
et al. 2000b, Ferrarese et al. 2001, McLure \& Dunlop 2002, Woo \& Urry
2002), provided correction factors related to the unknown geometry of
the broad emission line region are used. 
 
The results discussed above imply that the formation and growth of the
nuclear black hole and the bulge in a galaxy are intimately related, and
several theoretical models have attempted to explain the observed \mbh
-- $\sigma$ and \mbh--L$_{Bulge}$ relations (e.g. Haehnelt 2003,
Haehnelt et al. 1998, Adams et al. 2001, King 2003).  It is of interest,
therefore, to follow the tracks of AGNs in the \mbh -- $\sigma$ or \mbh
-- L$_{Bulge}$ plane to discriminate among models and so to understand
this concurrent growth.  Since high accretion rates would lead to
substantial black hole growth, active galaxies with close to Eddington
accretion are perhaps the best candidates. At low redshift, abundant
observational evidence suggests that narrow line Seyfert 1 galaxies
(NLS1s; a subclass of Seyfert galaxies, whose most notable feature is
that the full widths at half maximum of H$\beta$ lines are less than
2000 \kms; Osterbrock \& Pogge 1985; Goodrich 1989) accrete at close to
Eddington rate (e.g., Pounds et al. 1995, Grupe 2004 and references
therein). NLS1s are found to lie below the \mbh--bulge relations (both
\mbh--\sig and \mbh--L$_{Bulge}$) of normal galaxies and broad line
Seyfert 1s (BLS1s) (Mathur et al. 2001, Czerny et al. 2001, Wandel
2002). Using a complete sample of soft X-ray selected AGNs, this result
was confirmed by Grupe \& Mathur (2004; see also Mathur \& Grupe 2005a,
2005b; Watson et al. 2007). The statistical result is robust and is not
due to any systematic measurement error. However, it was obtained by
using FWHM([OIII]) as a surrogate for the bulge velocity dispersion and
the black hole mass was estimated using the optical (5100\AA)
luminosity, FWHM(H$\beta$), and the scaling relations (e.g. Bentz et
al. 2006). Black hole mass estimates using two other methods gave
consistent results, suggesting that the BH masses are not
underestimated; Mathur et al. (2001) used accretion disk model fits to
estimate BH masses, and Nikolajuk et al. (2009) used variability power
density spectra for the purpose. However, the use of FWHM([OIII]) as a
surrogate for velocity dispersion was questioned by many authors. For
example, Komossa \& Xu (2007) argued that if the [OIII] line is
blueshifted, or if it has asymmetric blue wings, then it ceases to be a
good indicator of the bulge velocity dispersion. Therefore, it is
necessary to confirm or refute whether NLS1s really lie below the
\mbh--\sig relation. Similarly, it would be useful to make accurate
measurements of L$_{Bulge}$ and confirm whether NLS1s lie below the
\mbh--L$_{Bulge}$ relation.  If confirmed, it would imply that at
low-redshift black holes grow by accretion in well-formed bulges; as
they grow, they may move closer to the \mbh--\sig\ or \mbh--L$_{Bulge}$
relations for normal galaxies.  This result, if confirmed, would not
support theories of \mbh--\sig\ relation in which the black hole mass is
a constant fraction of the bulge mass {\it at all times} in the life of
an AGN or those in which bulge growth is controlled by AGN feedback (as
discussed in detail by Mathur \& Grupe 2005b). If, on the other hand, we
find that the NLS1s do lie on the \mbh-bulge relations, the implications
are again interesting. The NLS1s in the sample of Grupe \& Mathur have
smaller mass BHs compared to BLS1s; if they all follow the same
\mbh-bulge relations then it would imply that in the present epoch,
highly accreting supermassive black holes exist in smaller
galaxies. This will provide important input to our understanding of
BH--galaxy co-evolution.

Using an optically selected sample of NLS1s (Williams et al. 2002),
Williams et al. (2004) have shown that NLS1 galaxies are a mixed
bag. While some show steep X-ray spectra and strong FeII emission, some
do not. Those NLS1s which show these extreme properties are highly
likely to be accreting at a high Eddington rate and those are the ones
which lie below the \mbh--\sig\ relation (Mathur \& Grupe 2005b). There
are undoubtedly some NLS1s having narrow emission lines as a result of
their face-on orientation. The newly discovered $\gamma$-ray active
NLS1s observed with $Fermi$ (Abdo et al. 2009) are most likely viewed
close to the jet axis, confirming earlier similar conclusions from radio
observations (Komossa et al. 2006, Yuan et al.  2008). However, as a
class, NLS1s are much less radio-loud than BLS1s, arguing against a
predominance of beaming and face-on views in NLS1s as a class (Komossa
et al. 2006). It is quite possible that these beamed NLS1s are not
highly accreting.  Our focus here, however, is on objects with evidence
for high Eddington accretion.

We performed HST/ACS imaging of host galaxies of a sample of 10 NLS1s
that appear not to lie on the \mbh--\sig\ relation in Grupe \& Mathur
(2004).  In principle, we could measure \sig\ directly with the CaII
triplet line.  However, for many of the NLS1s in our sample, the CaII
lines fall in the water vapor band in the Earth's atmosphere. In many
NLS1s for which CaII line is accessible from ground, CaII is observed in
emission rather than in absorption (Rodriguez-Ardila et al. 2002). This
makes the use of stellar absorption features to determine \sig\
difficult for the targets of interest. The goal of the HST observations
was to measure the fundamental plane parameters $\langle I_e\rangle$ and
$R_e$ and so get an alternative handle on \sig (the bulge luminosity is
related to the stellar velocity dispersion $\sigma_*$ through the
fundamental plane of galaxies with log R$_e= 1.24 \log\sigma -0.82 \log
\langle I_e\rangle -$C where R$_e$ is the effective radius of the bulge
profile and $\langle I_e\rangle$ is the average surface brightness
inside R$_e$ (e.g.  Dressler et al. 1987; Djorgovski \& Davis 1987)). We
could then plot the loci of our galaxies on both the \mbh--L$_{Bulge}$
and \mbh--\sig planes. However, we show in \S 4 that the bulges of at
least some of our target galaxies are actually pseudobulges which do not
lie on the fundamental plane. As such, we cannot infer \sig and so
cannot determine the locus of our galaxies on the \mbh--\sig plane. The
HST observations, however, allowed us make accurate measurements of
bulge luminosity and so find the locus of our galaxies on the
\mbh--L$_{Bulge}$ plane (\S 4).

The layout of the paper is as follows. In section 2 we present
observations and data reduction. The analysis is presented in section 3
and the results in section 4. We then present a comprehensive discussion
in section 5 and conclude in section 6.


\section{Observations \& Data Reduction}

\subsection{Observations}

 We observed with the ACS/HRC a sample of 10 NLS1s that do not appear to
 lie on the \mbh--\sig\ relation (selected from Grupe \& Mathur 2004)
 and are close enough to achieve deconvolution of the bulge and the
 AGN. The details of the observations are given in Table 1.  In Table 2
 we present optical and X-ray properties of the sample galaxies that are
 generally used to define the NLS1 class; these are the FWHM(H$\beta$),
 the X-ray power-law slope $\alpha_X$, the ratio of equivalent widths of
 FeII to H$\beta$, and the Eddington luminosity ratio. We also list the
 BH mass estimates from Grupe \& Mathur (2004). In addition we quote the
 mean values of all the parameters with dispersion for the complete
 sample of soft X-ray selected NLS1s from Grupe et al. (2004). It can be
 clearly seen that the 10 NLS1s discussed in this paper are
 representative of the NLS1 population in general.

We used the F625W filter taking advantage of the fact that the HRC pixel
 size critically samples the PSF at 6300 \AA.  The large luminosity of
 the central AGN makes detecting the faint host galaxy underneath it
 difficult.  To expand the dynamic range of the HRC, we took multiple
 exposures of each Seyfert galaxy: short exposures to well-measure the
 central core of the AGN's PSF, and longer exposures to pull out the
 surface brightness of the bulge (and disk) of the host galaxy.

\subsection{Data Reduction}

We used the standard ACS pipeline to reduce the data, with STSDAS in
IRAF\footnote{http://iraf.noao.edu/}.  However, reduction of our
observations was complicated due to the fact that all galaxies have a
bright nucleus. This made it necessary to take several steps which are
beyond the standard ACS pipeline\footnote{see http://www.stsci.edu/hst/acs/documents/handbooks/currentDHB/acs$_{-}$cover.html
for details on ACS data reduction}.

\noindent
{\it Pixel Repair:} 
   In half of our sample, one or more pixels in the core of the central
   AGN's PSF were saturated.  In addition, electrons can spill over into
   pixels around a saturated pixel; this occurred most commonly to
   pixels in the same column as the saturated pixel, most likely during
   read-out.  We therefore had to ``repair'' these pixels by replacing
   the value of its corrupted data number with a data number from a
   shorter exposure scaled by the exposure lengths.  This process proved
   satisfactory for four of the five AGNs with saturated pixels, but in
   the case of IRASF12397 we found that even the shortest exposure was
   saturated in the very central pixel. However, a basic Gaussian fit
   indicated that any bleeding along the column was minimal, so we left
   this single pixel unrepaired in all of the exposures. 
The contribution of the pixel repair to the
   total profile uncertainty was found to be negligible compared to
   the much larger issue of the change in instrument focus and the
   effects that had on the PSF core/wings ratio (see below).

\noindent
{\it Cosmic Ray Rejection:} Utilizing the same method as Bentz et
   al. (2009), we used the cosmic-ray rejection routine 'lacosmic' (van
   Dokkum 2001) to create a cosmic-ray mask of each of our images.  We
   then removed any cosmic-ray flag from the central 11x11 pixels as
   well as removing any flags associated with the Airy ``ring''.  The
   PSF of the HRC does not produce a smooth Airy ring but instead
   produces a constellation of light ``bulges'' in place of a ring.
   This complicates cosmic-ray rejection algorithms and so we took
   manual control in this process to prevent the light in the
   constellated Airy ``ring'' from being damaged by the cosmic ray
   rejection routines.  To ensure that no cosmic-ray had landed in the
   core of the AGN PSFs, we used a radial profile function (in {\it
   imexamine} within STSDAS) to look for discrepancies and then manually
   repaired affected pixels.  The cosmic-ray mask was converted into a
   cosmic-ray list and the final cosmic-ray repair (both automated and
   manual) was handled in
   XVista\footnote{http://ganymede.nmsu.edu/holtz/xvista/}.

\noindent
{\it Registration:} To combine our images as accurately as possible, we
   needed registration at the sub-pixel scale.  We did this by
   centroiding the light from the AGN PSF of each exposure.  The optimal
   method would involve averaging multiple point-sources across the chip
   and avoiding any extended source.  However, we lacked any stars in
   the field of view of the majority of our targets.  We found that
   centroiding only the AGN light was actually quite robust, for while
   there is indeed an extended source around it (the host galaxy), it is
   so faint that it does not affect the process.  We used both {\it
   qphot} and {\it imexamine} to determine the centroid of the AGN PSF.
   In the majority of cases these two routines found the same centroid
   to within 0.05 of a pixel.
   Overall, the corrections to the preprogrammed dithering were under
   0.07 pixels but could vary from an average of 0.04 pixels for TONS180
   to 0.17 pixels for MRK478.  The images then were coadded and the
   cosmic rays were removed, but did not (yet) undergo geometric
   undistortion.

We found that the library PSFs available for the ACS/HRC 
 were insufficient in matching the nuclear light. Therefore we created a PSF
   template from a bright star in the field of RXJ2217 (at a projected
   distance of 36\,kpc so we are confident of no contamination) and
   performed a similar procedure to that discussed above.  This star was
   (barely) unsaturated in the shortest exposures and we repaired the
   central pixels in the longer exposures.  We then pulled a 201x201
   pixel square around the star out of the image, detected and repaired
   the pixels hit by cosmic rays (of which there were few and none in
   the central core or Airy ``ring'') using {\it lacosmic} and XVista.
   We then fit the centroid of the PSF, and {\it imshift} to align the
   eight individual images, and coadded them in XVista.  Finally, as the
   algorithm we used to deconvolve the components of this image (GALFIT;
   Peng et al. 2002) requires a sky-subtracted reference PSF, we fit the
   sky in the corners of the image away from the PSF halo and the PSF
   diffraction spikes and then subtracted the average scalar value from
   the image.

 Finally, we fit each coadded (but not yet corrected for the geometric
   distortion of the HRC) image with this stellar PSF (representing the
   nuclear source), a deVaucouleurs profile (representing the central
   bulge), an exponential profile (representing the underlying disk),
   and a background sky.  The purpose at this time was not to achieve a
   match to the galactic profile but rather to the nuclear source.  In
   many cases manual control over the GALFIT routine had to be taken in
   order to provide the best match to the nuclear source.  Because
   temperature fluctuations across HST change the focus of the
   instrument slightly (``breathing''), matches to both the core of the
   PSF and the outer regions was never perfect as each object was
   observed for one orbit.  But as our purpose in this project is to
   best model the underlying galaxy, and these galaxies are relatively
   close, we concentrated upon matching the outer portions of the PSF.
   The metric used was the number of artifacts from the Airy ring and
   the diffraction spikes that remained upon subtraction of the PSF.
   Because the Airy ring in ACS/HRC has a beaded structure, it was
   rather easy to test for proper registration, PSF modeling, and PSF
   subtraction; a smoothed profile resulted in nine of the ten objects
   studied (please see figures 1 and 2 for the residuals of TONS180 for
   evidence of where our constructed PSF did not match as well as the
   other nine of our sample.  TONS180 also has the brightest AGN in our
   sample which magnified this issue).  The other major source of our
   ability to match the nuclear PSF was that this was all done before
   the correction for the geometric distortion of the HRC.
   As the geometric correction interpolates pixel
   positions, it naturally distorts the intrinsic PSF.  The resultant
   PSF-subtracted images were then put back into PyDrizzle for the
   geometric correction to give us the final images from which to fit
   the galactic profiles (Figure 1).

\section{Analysis}

Using GALFIT, all ten targets were initially fit using three profiles: a
fixed sky (previously derived from the corners of the image), and Sersic
profiles, one started with $n=3.5$ and $R_e=1.0$ kpc (representing the
bulge), and one with $n=1$ (fixed) and starting with $R_e=3.0$ kpc
(representing the disk).  An iterative process of examining the
residuals and adding additional components was done for each target,
testing each for significance (this also sometimes included data
reduction components like residual PSFs or outright masking of the
pixels associated with the PSF core for those galaxies whose PSF
core/wings ratio was not as well fit by our model PSF).  The expected
features were found amongst our targets: non-traditional profiles,
spiral arms, stellar bars, dust lanes, and (projected) nearby
companions.  In half the sample nearby companions (projected distance
$\sim 10$kpc and greater) were discovered but as these were not the
focus of this study, the regions they occupied were masked out (in only
one object did the companion overlap the visual extent of the target).
Only one object was visually affected by asymmetric crossing dust lanes,
but as these lanes were not easily excised by masking, manual control
was taken of the fitting and the lanes were taken as residuals.  Stellar
bars of varying sizes were found (to varying degrees of certainty) in
approximately 1/3 to 1/2 of the sample and the brightest were fit using
a very boxy, high axis ratio profile with the proper position angle;
please see the the discussion in \S 4 and the residuals column in Figure
1 for details.  Spiral arms were generally taken into account by manual
control in the last stages of the fitting.  In fact, in all ten cases
manual control was used to provide a comparison against the best fit
made by the fitting algorithm to ensure against erroneous local minima.

The best fit profile parameters of all the targets are presented in
Table 2.  The bulges achieve an azimuthally integrated S/N ratio at
their $R_e$ between 150 to 360. The final profile fits are shown in
figures 3 \& 4; the black line shows the data while the dotted and
dashed blue lines show the disk and bulge components respectively. The
red line is the sum of all components and shows that the bulge$+$disk
profile fits the data well. There are small imperfections in two cases:
RXJ$2216.8-4451$ and RXJ$1702.5+3247$.  The galaxy RXJ$2216.8-4451$ is
an asymmetrically dusty galaxy.  The dust does not fall in lanes but
rather in an irregular-shaped ring around the center at a distance of
$\sim2$ kpc (as seen in fig. 1).  This results in the temporary
flattening seen in its the radial profile from about 1.5 to 2.5 kpc
(Fig. 3).  The galaxy RXJ$1702.5+3247$ is our furthest galaxy and also
the one with the poorest initial deconvolution (\S2.2).  As can be seen
in fig.4, there are significant artifacts left over from PSF subtraction
(the two peaks correspond to the core and first Airy ring, as seen in
fig. 2).  The most likely cause are thermal fluctuations across the
orbit (``breathing'').  The disk profile is clearly seen but because the
bulge profile may still be contaminated by the wings of the PSF, we
caution that the reported bulge parameters of this galaxy are our most
uncertain. The galaxy RX\,J1209.8+3247 only contains one component with
$n=1$. This may be interpreted as a bulge-less disk or a pseudobulge
with $n=1$ (see \S 4) and a fainter, undetected extended disk (see \S 4
for a discussion on pseudobulges). The presence of a supermassive black
hole in this galaxy would be perhaps more surprising (and interesting)
than the presence of a pseudobulge; pseudobulges with $n=1$ have been
observed previously (Kormendy \& Kennicutt 2004 (KK04)) and so are AGNs
in bulge-less galaxies (Araya et al. 2012).

One additional series of tests was done to ensure the quality of the
fits.  While in all cases but one, a bulge-like component and a
disk-like component were found to be quite significant, comparisons were
done between fits which (1) let the Sersic profiles of the bulge
component entirely unconstrained, (2) limited the Sersic profile to $1
\leq n \leq 4$ and (3) fixed the Sersic profiles at $n=4$.  Masking was
also done on the central pixels which were within the limits of the core
of the nuclear PSF as well to test the profile fitting.  In the majority
of cases, GALFIT naturally found Sersic profiles between 1.0 and 4.0.
We discovered that five of our targets naturally had a low ($n$ \lax
$2.2$) Sersic-profile small-scale component. One interpretation of this
result is that these five galaxies have pseudobulges; we shall come back
to whether these, and other galaxies with $n > 2$, are real pseudobulges
in \S4.  Of the remaining five, GALFIT found a Sersic index higher than
$n=4.0$ in two cases. The other two were constrained to $n=4.0$, because
GALFIT fit did not converge close to $n=4$. If left unconstrained, the
``bulge'' profile became steep and narrow, indistinguishable from a
point source. One way of interpreting this might be that in these
galaxies only one ($n=1$) component is required and this may be a $n=1$
(pseudo)bulge with a faint. undetected disk, similar to what is observed
for RX\,J1209.8+3247. This will increase our pseudobulge candidates from
5 to 7; here we take a conservative approach and constrain $n=4$.

In general the sizes of the bulges we found are less than 2.0 kpc (a
cosmology of $\Omega_m=0.3, \Omega_{\Lambda}=0.7$ was used to convert
arcsec to kpc, but all targets are relatively close by, at $z< 0.17$).
While the range of effective radii of bulges is within the normal range
observed, we need to examine whether they are smaller than the true
values, perhaps because of poor PSF subtraction.  If this were the case
then one would expect that the closest galaxies would have the
"smallest" bulges while the furthest galaxies have the largest bulges
because the PSF is a constant number of pixels but the plate scale
changes by a factor of four.  Such is not the case.  In addition, the
ratio of bulge effective radius to the visual extent of the PSF core has a
median of $\sim 3$ (smallest two are $0.9$ and $1.3$ while the rest are
larger than $2$).  So from this test, our bulge sizes appear to be real.
Note also that Gadotti (2009) has shown that the structural properties
of bulges can be reliably retrieved provided that the effective radius is
larger than about 80\% of the PSF HWHM; our measurements are
significantly above this minimum requirement. The pseudobulge candidates
(i.e., those with $n$ \lax $2$) ranged from 0.27 to 1.52 kpc in size and
the classical bulges (i.e., those with $n > 2.5$) ranged from 0.37 to
1.34 kpc.  Moreover, excess nuclear light left in the image would make
the Sersic profile steeper, but the galaxies with the smallest ratio of
$R_e$ to $R_{PSF}$ tended to be pseudobulge candidates instead
(i.e. with flatter profiles). We thus conclude that our measurements of
$R_e$ and $n$ do not suffer from artifacts of PSF subtraction.

Like the effective radii, the luminosities of these bulges also span
more than a factor of ten.  The images of these galaxies were taken in the
F625W filter, more commonly known as SDSS $r'$.  The absolute $r'$
magnitude of these bulges ranged from $-19.1$ mag to $-21.2$ mag with a
median of about $-20.1$ mag.  The five galaxies with the pseudobulge
candidates also had the four least-luminous bulges in our sample, again
consistent with the $L$--$n$ relation. One consistency check would be 
 to compare the HST and SDSS results for 
galaxies that have SDSS imaging.  In those five cases, the
absolute $r'$ magnitude of our bulges were always less than the total
absolute $r'$ magnitude detected by SDSS, as expected.


 Our pseudobulge candidates do have properties consistent with the
findings of other studies.  Fisher \& Drory (2008) have shown that while
classical bulges follow the structural parameters and the photometric
projections of the fundamental plane, pseudobulges do not. While there
is clear overlap in the structural parameters of bulges and pseudobulges
in their sample, we note that the pseudobulge candidates of our sample
occupy the same parameter space as that of pseudobulges of Fisher \&
Drory (their figure 7); this provides a good consistency check.  We have
also tested whether redshift could be responsible for our
low-Sersic-index bulges. While two of our pseudobulge candidates do have
our lowest ratio of bulge effective radius to PSF core, the remainder
all have ratios $\geq 2.5$. Among the pseudobulge candidates, one is the
lowest redshift source and four are at high redshift, but two similarly
high redshift sources also have high $n$ values. Thus our fitted
parameters do not appear to be spurious on account of redshift.

\section{Results}

As pointed out by KK04, a pseudobulge is not an observational
classification; it is a bulge made out of disk material by secular
evolution. It is difficult to determine whether a bulge in a galaxy is a
classical bulge (merger-driven) or a pseudobulge (secularly formed), but
there are several indicators. Pseudobulges have one or more
characteristics of disks: (1) flatter shapes; (2) larger ratios of
ordered to random velocities; (3) small velocity dispersion \sig\ with
respect to the Faber-Jackson relation between \sig\ and bulge
luminosity; (4) spiral structure or nuclear bars in the "bulge" part of
the light profile; (5) S\'{e}rsic index $n$ of the bulge surface
brightness profile \lax 2; and (6) dominance of population I material
(young stars, gas and dust) without a sign of a merger in progress
(KK04). Pseudobulges do not follow the Kormendy relation (Kormendy 1977;
a relation between surface brightness and effective radius);
pseudobulges are fainter at a fixed effective radius (e.g. Fisher \&
Drory 2008; Gadotti 2009).

We find that five of the ten galaxies have Sersic indices consistent with
 pseudobulges.  All the pseudobulges in Gadotti (2008),
 irrespective of the Sersic index, have small bulge scale lengths with
 $R_e$\lax $2$kpc; all our galaxies have similarly small bulges. It is
 thus possible that some of our classical bulge candidates are also
 pseudobulges. 


 


We can now find out where the NLS1 galaxies in our sample belong in
terms of bulge properties and black hole mass.  For the black hole mass
estimate we use the standard single-epoch procedure of using the
$L_{5100\AA}$ and FWHM(H$\beta$) of these NLS1s (see Grupe \& Mathur
2004 for details). The most straightforward approach is to compare the
black hole mass against the bulge luminosity, the so-called ``Magorrian
relation'' (Magorrian et al. 1998).  The most recent update on the
Magorrian relation is published by G\"{u}ltekin et al. (2009) who have
determined bulge luminosity in the $V$ band. Our observations, however,
are in the $r$-band, so we performed following color corrections.  We
use the average relation from Jester et al. (2005): $V = g - 0.585 *
(g-r) - 0.01$. To obtain the $g-r$ color corrections we considered two
options.  One is to consider the bulge a part of the red sequence; we
adopt the color sequence found in Bernadi et al. (2003) of $g-r = 0.218
- 0.025 M_r$ (which gives an average color of $\sim 0.73$) for all our
bulges with $n > 1.5$.  Pseudobulges, however, have been found to be
still actively forming stars (KK04), so the $g-r$
color likely is more indicative of an average blue cloud galaxy,
i.e. $g-r =0.5$.  We show the result in Figure 5 taking into account
both these color corrections; it is clear that the results are
independent of the choice of color corrections.  The figure shows that
our NLS1s do not lie on the ``Magorrian relation'', but are actually
below it. This confirms the results of Mathur and collaborators
discussed in \S1.  Earlier we could not determine whether NLS1 bulges
are generally overluminous for their black holes, or NLS1 black holes
are undermassive for their bulges. We can rule out the former for at
least part of our sample with pseudobulges; for the fixed size,
pseudobulges are typically less luminous than classical bulges (Gadotti,
2009; Fisher \& Drory 2010). Thus it appears that the black holes in
NLS1s are undermassive for their bulges.

We also compared our data with the ``Magorrian relation'' for broad-line
AGNs (BLAGN) by Shen et al. (2008) and found consistent result, viz. our
NLS1s do not follow the \mbh--L$_{bulge}$ relation of BLAGN.  It should
be noted that the Shen et al. luminosities are for the entire host
galaxy (not just the bulge); taking this into account enhances the
difference between NLS1s and BLAGN.

We have to keep in mind that our results may be affected by dust. As
noted above our observations are in the HST $r$-band only, so we do not
have any color information available for our sample. Naturally, the
amount of extinction and reddening depends on the inclination of the
galaxy, with face-on galaxies least affected. While one may determine
the inclination for a disk by measuring the axial ratio, the same cannot
be done for bulges, because we do not know their shape a priori. While it
is not known whether the bulge and the disk of a galaxy have the same
rotation axis, we can assume them to be the same for the time being. In
Table 2 we have listed the axial ratios of the disks of our galaxies;
most of them are pretty close to 1, indicating they are close to face
on, so the effect of dust is minimal. Graham \& Worley (2008) have
given prescription for correcting the observed bulge magnitude for dust
and for correcting the observed disk scale height and the disk central
surface brightness; we will assume the same corrections for the bulge as
well.  Accordingly, our data points in Fig. 5 will move to right (higher
luminosity) by a minimum of $0.56$ (and more for more inclined
galaxies), strengthening the result.

\section{Discussion}

HST/ACS observations of our sample of 10 NLS1 galaxies have revealed
that the bulge profiles range from ``classical'',
i.e. near-deVaucouleurs, for five galaxies to ``pseudo'',
i.e. near-exponential, for five galaxies. 
Thus at least half of our sample galaxies likely host pseudobulges. This
is consistent with the results of Orban de Xivry et al. (2011) who find
that NLS1s hosts preferentially have pseudobulges while BLS1 hosts
preferentially have classical bulges.

Hu (2008) has shown that the \mbh--\sig\ relation for pseudobulges is
different from the relation for early-type bulges.  Gadotti \& Kauffmann
(2009) have also shown that the BHs in pseudobulges do not follow the
\mbh--\sig\ relation of normal galaxies (and broad-line AGNs); they lie
below the relation. Indeed, we find that the galaxies in our sample also
lie below \mbh--bulge luminosity relation. (We did not find the locus of
our galaxies on the \mbh--\sig\ plane, because we do not have direct
measurements of \sig; pseudobulges do not follow the fundamental plane
either, so we cannot use the observed $<I_e>$ and $R_e$ measurements to
estimate \sig). These results are fully consistent with the earlier
results of Mathur \& collaborators (\S 1) which showed that the locus of
NLS1s on \mbh--bulge relations is distinct from that of BLS1s and normal
galaxies. Neither the black hole mass estimates nor the estimates of
\sig\ based on narrow-line widths give a spurious result. Thus the black
holes in NLS1 galaxies are truly undermassive for their bulges. They
are, however, growing at a close-to-Eddington rate, so may reach the
scaling relations of BLS1s eventually (Mathur 2000), provided they
continue to accrete at the present rate. On the other hand, they may
never reach the BLS1 scaling relations, especially if their BHs are
growing slowly (Orban de Xivry et al. 2011).

As noted above, Gadotti (2008) has shown that pseudobulges do not lie on
the fundamental plane. We can check whether our sample galaxies lie on
the ``photometric plane'' defined by Graham (2002). In figure 6, we have
plotted the effective radius $\log(R_e)$ vs. $\log(n)+ b<I_e>$ where
$b=0.26$, the photometric plane. Points are our data and the solid line
is the correlation found by Graham (2002) for E and S0 galaxies. The
scatter around the correlation was found to be about $0.125$ dex in
$\log R_e$ (dotted lines in Fig. 6). We see that four of our galaxies
are off the line on this hyperplane, with offsets much larger than the
scatter. Thus it seems that the pseudobulges do not lie on the
photometric plane either.  We note, however, that our observations are
in the $r$-band, while the photometric plane is defined in the
B-band. The color correction will move the points to the right by about
0.052, which will not affect our conclusion. Moreover, La Barbera et
al. (2005) have shown that the photometric plane relation is independent
of the waveband. Taking into account dust correction, the data points
will move down (lower $R_e$) by a 0.045 and to the left (lower
$<\mu>_e$) by 0.06, effectively moving slightly away from the line,
again strengthening the result. We also find that all 10 objects in our
sample are offset from the projection of the fundamental plane of Barway
\& Kembhavi (2007). Given that our sample is of late-type galaxies,
perhaps a comparison with the photometric plane of dwarf ellipticals
(Kourkchi et al. 2011) is more appropriate. We again find that our
pseudobulge candidates lie off the relationship presented in Kourkchi et
al.

Black hole masses in Seyfert galaxies are measured through reverberation
mapping (Peterson 1993) when possible. The unknown geometry of the broad
emission line region, however, leads to an uncertainty of the order of
0.5dex in measured masses, which is characterized by a factor
$f$. Collin et al. (2006) estimated the value of $f$ under the assumption
that all black holes lie on the \mbh--\sig\ relation, and the observed
scatter is solely due to the scatter in $f$. Because NLS1s were found to
lie systematically below the \mbh--\sig\ relation, they derived a higher
value of $f$ for NLS1s, compared to BLS1s, to move them back on the
\mbh--\sig\ relation. The results presented in this paper show that the
Collin et al. result is not really valid for NLS1s because they occupy
an intrinsically different locus on the \mbh--\sig\ plane because their
bulges are different. At this point it appears that calibrating the AGN
mass scale with the \mbh--bulge relationships (with \sig or $L$) will be
either more difficult or more uncertain than hoped.

The \mbh--bulge relations (with \sig or $L$) are also used to determine
black hole masses in sources in which broad emission lines are not
easily observed, e.g., blazars (Falomo et al. 2002), radio galaxies
(Bettoni et al. 2003), bright cluster galaxies (Lauer et al. 2007;
Batcheldor et al. 2007), and obscured AGNs (e.g. Greene et
al. 2009). McLure \& Dunlop (2002) have also used the scaling relations
to determine black hole masses of AGNs and suggest their use to
determine black hole masses of high redshift galaxies.  Because of the
intrinsic scatter in these relations caused by a combination of factors
such as the galaxy morphology, orientation and redshift evolution,
application of scaling relations to determine BH masses may not be as
straightforward as has been hoped.

As noted above, at least five of our NLS1 host galaxies have low-n bulges. 
If they are true pseudobulges, it has implications for black
hole growth. Pseudobulges tend to show younger stellar populations as
well as distinct structural and kinematic properties (rotational
support), indicating different formation processes. While classical
bulges are believed to have formed through mergers, pseudobulges are
likely formed through secular evolution or long distance
interactions. The pseudobulges are perhaps still in the formation
process. It then follows that their nuclear supermassive black holes are
also recently formed, and still growing. This is exactly what we had
proposed for NLS1s, because they accrete at a higher Eddington rate
compared to BLS1s (Mathur 2000). Thus the youth of NLS1s is supported by
their growing black holes as well as their pseudobulge hosts. These
results also suggest that there are different modes of black hole
growth. At high redshift, the black holes appear to have grown quickly
through merger-driven processes. These BHs are massive and inactive at
the present epoch and reside in the centers of elliptical galaxies. In
the galaxies hosting pseudobulges, the black holes are in the growth
mode at the present epoch and the growth is triggered by secular
processes.  A secular slow, long, period of BH growth is also possible
(Orban de Xivry et al. 2011). It should also be noted that the
alternative, secular, mode of black hole growth is perhaps a dominant
one at the present epoch. Weinzirl et al. (2009) have shown that about
70--75\% of high-mass spirals contain pseudobulges, based on the values
of Sersic index or the B/T ratio; Fisher \& Drory (2011) have also come
to similar conclusion. Given that spirals outnumber ellipticals, it
follows that the growth of black holes in most galaxies follows a
secular track.

\section{Conclusions}

We observed ten NLS1 host galaxies with HST/ACS.  We find that our
 sample AGNs lie below the ``Magorrian relation'' of normal galaxies and
 BLS1s, confirming earlier results of Mathur et al. We caution against
 using the \mbh--bulge relations to determine black hole masses or to
 determine the geometry of the broad line region of AGNs. Image analysis
 revealed that five of them likely host pseudobulges. If they are true
 pseudobulges, it would imply that secular processes play important roles
 in galaxy evolution and black hole growth and that this alternative
 track of black hole--galaxy evolution may in fact be a dominant one.

Acknowledgment: This work is supported in part by the STScI grant
HST-GO-10436 to SM and by the NSF grant AST-1008882 to BMP.

=========================================================================



\newpage
References:

Abdo, A.A., et al. 2009, ApJ, 707L, 142

Adams, F.C., Graff, D.S., \& Richstone, D.O, 2001, ApJ, 551, L31 

Aller, M.C., \& Richstone, D., 2002, AJ, 124, 3035 

Barger, A.J., Cowie, L.L., Bautz, M.W., Brandt, W.N., et al., 2001, AJ,
122, 2177 

Barway, S. \& Kembhavi, A. 2007, ApJ, 662, L67

Batcheldor, D., Marconi, A., Merritt, D., \& Axon, D., 2007, ApJ, 663, L85 

Bentz, M. et al. 2006, ApJ, 644, 133

Bentz, M. et al. 2009, ApJ, 697, 160

Bernadi et al. 2003 ApJ 125 1882

Bettoni, D., Falomo, R., Fasano, G., \& Govani, F. 2003, A\&A, 399, 869

Collin, S., Kawagichi, T., Peterson, B.M., Vestergaard, M. 2006, A\&A, 456, 75

Czerny, B. et al. 2001, MNRAS, 325, 865 

Falomo, R., Kotilamen, J.K., \& Treves, A., 2002, ApJ, 569, L35

Ferrarese, L., \& Merritt, D., 2000, ApJ, 539, L9 

Ferrarese, L., Pogge, R.W., Peterson,
B.M., Merritt, D., Wandel, A., \& Joseph, C.L., 2001, ApJ, 555, L55 

Fisher, D. B., \& Drory, N., 2008, AJ, 136, 773

Gadotti, D. A. 2009, MNRAS, 393, 1531

Gadotti, D. A., \& Kauffmann, G. 2009, MNRAS, 399, 621

Gebhardt, K., Bender, R., Bower, G., Dressler, A., Faber, S.M., et al.,
2000a, A\&A, 539, L13

Gebhardt, K., Kormendy, J., Ho, L.C., Bender, R., Bower, G., et al.,
2000b, ApJ, 543, L5

Graham, A. 2001,  AJ, 121, 820

Graham, A. 2002, MNRAS, 334, 859

Graham, A., Onken, C. A., Athanassoula, E., \& Combes, F. 2010, arXiv:1007.3834

Grupe, D., 2004, AJ, in press (April 2004), astro-ph/0401167 

Grupe D. \& Mathur, S. 2004, ApJL, 606, 41

Greene, J., Zakamska, N., Liu, X., Barth, A., \& Ho, L. 2009, ApJ, 702, 441

G\"{u}ltekin, K., et al. 2009, ApJ, 698, 198

Haehnelt, M., 2003,  Classical and Quantum Gravity, 20, S31 

Haehnelt, M.G., Natarajan, P., \& Rees, M.J., 1998, MNRAS, 300, 817 

H\"{a}ring, N. \& Rix, H.W. 2004, ApJ, 604L, 89

Hu, J., 2008, MNRAS, 386, 2242

Jester, S., et al. 2005, AJ, 130, 873

Jorgensen, I. et al. 1996, MNRAS, 280, 167 

King, A., 2003, ApJL, 596, L27 

Komossa, S. et al. 2006, AJ, 132, 531

Komossa, S. \& Xu, D., 2007, 667L, 33

Kormendy, J., 1977, ApJ, 218, 333

Kormendy, J. \& Kennicutt, R. C. 2004, ARAA, 42, 603

Kourkchi, E., Khosroshahi, H.G., Carter, D. \& Mobasher, B. 2011, MNRAS,
420, 2835

La Barbera, F. et al. 2004, A\&A, 425, 797

La Barbera, F. et al. 2008, ApJ, 689, 913


Lauer, T.R. et al. 2007, ApJ, 662, 808 

Magorrian, I. et al. 1998 , AJ, 115, 2285 

Mathur, S., 2000a, MNRAS, 314L, 17

Mathur, S. 2000b, NewAR, 44, 469

Mathur, S., Kuraszkiewicz, J., \& Czerny, B., 2001, New Astronomy,
Vol. 6, p321

Mathur, S. \& Grupe D. 2005a, ApJ, 633, 688

Mathur, S. \& Grupe D. 2005b, A\&A, 432, 463

McLure, R.J., \& Dunlop, J.S., 2002, MNRAS, 331, 795 

Merritt, D., \& Ferrarese, L., 2001, ApJ, 547, 140

Nikolajuk, M., Czerny, B., \& Gurynowicz, P., 2009, MNRAS, 394, 2141 

Orban de Xivry et al. 2011, in preparation. 

Osterbrock, D.E., \& Pogge, R.W., 1985, ApJ, 297, 166 

Peng, C. Y., Ho, L. C., Impey, C. D. \& Rix, H. 2002, AJ, 124, 266

Peterson, B.M. 1993, PASP, 105, 247

Pounds, K.A., Done, C., \& Osborne, J., 1995, MNRAS,277, L5 

Rodriguez-Ardila, A., Viegas, S.M., Pastoriza, M.G., \& Prato, L., 2002,
ApJ, 565, 140

Shen, Van den Berk, Schneider, \& Hall 2008, AJ, 135 928

Tremaine, S., Gebhardt, K., Bender, R., et al., 2003, ApJ, 574, 740 

van Dokkum, P.G. 2001, PASP, 113, 1420

Wandel, A., 2002, ApJ, 565, 762 

Watson, L., Mathur, S. \& Grupe, D. 2007, AJ, 133, 2435

Williams, R., Pogge, R.W., \& Mathur, S. 2002, AJ, 124, 3042

Williams, R., Mathur, S., \& Pogge, R.W. 2004, ApJ, 610, 737

Weinzirl, T., et al. 2009, ApJ, 696, 411

Woo, J. \& Urry, C.M. 2002, ApJ, 579, 530 

Yu, Q., \& Tremaine, S., 2002, MNRAS, 335, 965

Yuan, W. et al. 2008, ApJ, 685, 801

\begin{deluxetable}{cccccc}
\tablecaption{Journal of Observations \label{tab:obs}}
\tablehead{
\colhead{Object}
&\colhead{Dataset}
&\colhead{Observation}
&\colhead{Exposure}
&\colhead{Redshift}
&\colhead{Scale} \\
\colhead{}
&\colhead{}
&\colhead{Date}
&\colhead{seconds}
&\colhead{}
&\colhead{[kpc/arcsec]} \\
}
\startdata
MS\,2254-36&J96I10&2005-04-15&2390&0.039&0.73 \\
IRAS\,F12397+3333&J96I09&2005-01-16&2464&0.044&0.84 \\
TONS\,180&J96I10&2005-08-18&1528&0.062&1.13 \\
MRK\,478&J96I02&2005-04-30&1542&0.077&1.44 \\
RXJ\,2216.8-4451&J96I03&2005-04-20&2428&0.136&2.31 \\
MS\,23409-1511&J96I04&2005-05-06&2150&0.137&2.32 \\
RXJ\,1209.8+3247&J96I05&2005-01-16&2558&0.145&2.46 \\
RXJ\,1117.1+6522&J96I06&2005-04-18&2544&0.147&2.48 \\
RXJ\,2217.9-5941&J96I07&2005-04-23&2542&0.160&2.65 \\
RXJ\,1702.5+3247&J96I08&2005-04-23&2370&0.164&2.70 \\
\enddata
\end{deluxetable}

\newpage
\small
\begin{deluxetable}{lccccccc}
\tablecaption{Properties of the AGN observed by HST in comparison with the mean
 values of NLS1s} 
\label{obj_list}
\tablewidth{0pt}
\tablehead{
\colhead{Object} 
& \colhead{$\alpha_{2000}$} 
&  \colhead{$\delta_{2000}$} 
& \colhead{FWHM(H$\beta$)\tablenotemark{1}}
& \colhead{$\alpha_X$\tablenotemark{2}}
& \colhead{Fe II/H$\beta$\tablenotemark{3}}
& \colhead{$L/L_{Edd}$\tablenotemark{4}}
& \colhead{$M_{\rm BH}$\tablenotemark{5}}
} 
\startdata
Ton S 180        & 00 57 20.2 & --22 22 57 & $970\pm100$  & 1.89 & 0.90 & 6.30 & 7.1 \\ 
RX J1117.1+6522  & 11 17 10.1 &  +65 22 07 & $1650\pm170$ & 1.89 & 0.99 & 0.40 & 21.0 \\
RX J1209.8+3217  & 12 09 45.2 &  +32 17 02 & $1320\pm110$ & 3.18 & 1.09 & 1.45 & 5.4 \\
IRAS 12397+3333  & 12 42 10.6 &  +33 17 03 & $1640\pm250$ & 2.02 & 1.79 & 0.76 & 4.5 \\
Mkn 478          & 14 42 07.5 &  +35 26 23 & $1630\pm150$ & 2.08 & 0.97 & 0.15 & 26.9 \\
RX J1702.5+3247  & 17 02 31.1 &  +32 47 20 & $1680\pm140$ & 2.13 & 0.98 & 1.86 & 21.7 \\
RX J2216.8--4451 & 22 16 53.2 & --44 51 57 & $1630\pm130$ & 2.48 & 1.13 & 1.78 & 16.7 \\
RX J2217.9--5941 & 22 17 56.6 & --59 41 30 & $1430\pm60$  & 2.69 & 0.96 & 1.02 & 12.4 \\
MS 2254-36       & 22 57 39.0 & --36 56 07 & $1530\pm120$ & 1.78 & 0.53 & 0.24 & 3.9 \\
MS 23409--1511   & 23 43 28.6 & --14 55 30 & $1030\pm100$ & 2.03 & 1.18 & 0.22 & 10.0 \\
Sample mean &               &           &    1451          &  2.2  & 1.05 & 1.4 & 12.96 \\
NLS1s mean     &    ---     &   ---      & 1380            & 1.96  & 0.99 & 1.79 & 23.2 \\
NLS1s dispersion  &    ---     &   ---      &   350      & 0.41 & 0.40 & 2.87 &  20.2 \\
\enddata

\tablenotetext{1}{FWHM(H$\beta$) given in units of km s$^{-1}$ as listed
in Grupe et al. 2004.} 
\tablenotetext{2}{$\alpha_X$ taken from the ROSAT
measurements as listed in Grupe et al. 2004.} 
\tablenotetext{3}{FeII/H$\beta$ ratio given in Grupe et al. 2004.}
\tablenotetext{4}{Eddington ratio $L/L_{Edd}$ derived from the Swift
observations discussed in Grupe et al. 2010.}  
\tablenotetext{5}{Black hole masses derived from the virial relation as
given by Kaspi et al. (2000). The values for the AGN here are in units
of $10^6$\msun, listed in Grupe et al. 2010..}
\end{deluxetable}

\newpage
\small
\begin{deluxetable}{cccccccccc}
\rotate
\tablecaption{Galaxy Properties\label{tab:prop}}
\tablehead{
\colhead{Object}
&\multicolumn{3}{c}{Bulge}
&\multicolumn{4}{c}{Disk}
&\colhead{log M$_{BH}$}
&\colhead{$<\mu>_{e,V}$} \\
\colhead{}
&\colhead{M$_r'$}
&\colhead{R$_e$ [\,kpc]}
&\colhead{n}
&\colhead{M$_r'$}
&\colhead{R$_e$ [\,kpc]}
&\colhead{n}
&\colhead{b/a}
&\colhead{(5100\,\AA)}
&\colhead{}
}
\startdata
MS\,2254-36&-19.1&0.27&2.12&-19.7&2.0&1&0.77&6.60&16.9  \\
IRASF\,12397+3333&-20.2&0.88&3.45&-19.8&3.6&1&0.42&6.67&18.4  \\
TON\,S180&-20.1&1.90&5.45&-20.1&4.5&1&0.85&6.85&20.2  \\
MRK\,478&-21.2&0.95&4.00\tablenotemark{a}&-21.2&5.9&1&0.96&7.44&17.7 \\
RX\,J2216.8-4451&-21.1&1.34&4.00\tablenotemark{a}&-21.1&3.85&1&0.94&7.23&18.8 \\
MS\,23409-1511&-20.7&0.37&5.31&-20.4&5.5&1&0.93&7.01& 16.1\\
RX\,J1209.8+3247&-19.8&1.12&1.00\tablenotemark{b}&...&...&...&...&6.75& 19.6 \\
RX\,J1117.1+6522&-19.7&0.45&0.62&-21.4&5.40&1&0.94&7.33&17.7  \\
RX\,J2217.9-5941&-19.6&0.33&0.78&-21.1&3.0&1&0.86&7.10& 17.2  \\
RX\,J1702.5+3247&-19.8&1.52&0.69\tablenotemark{c}&-20.8&6.7&1&0.91&7.34& 20.3\\
\enddata
\tablenotetext{a}
{Forced Sersic profile $1\leq$n$\leq4$}
\tablenotetext{b}
{Only one component necessary}
\tablenotetext{c}
{Detection of Bulge marginal}
\end{deluxetable}

\newpage

\begin{figure}
\psfig{file=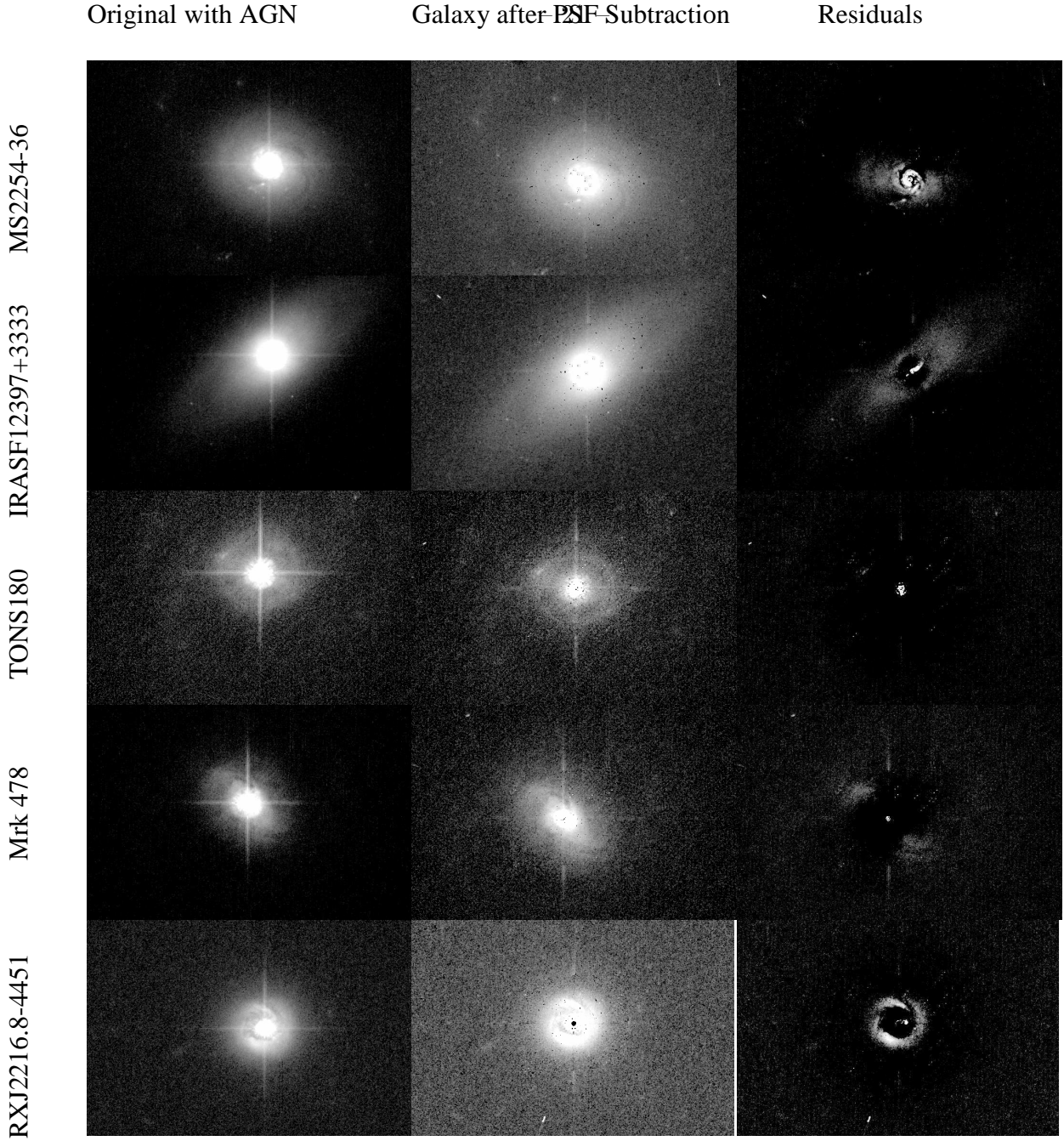,width=5in}
\caption{
ACS images of our sample galaxies. The left column shows the
observed images, and the middle column shows images after nuclear point
source subtraction. These images were fitted with bulge and disk
profiles; the residuals to the fit are shown in the right
column. Several images show structures such as bars and spiral
arms (see text).
}
\end{figure}

\begin{figure}
\psfig{file=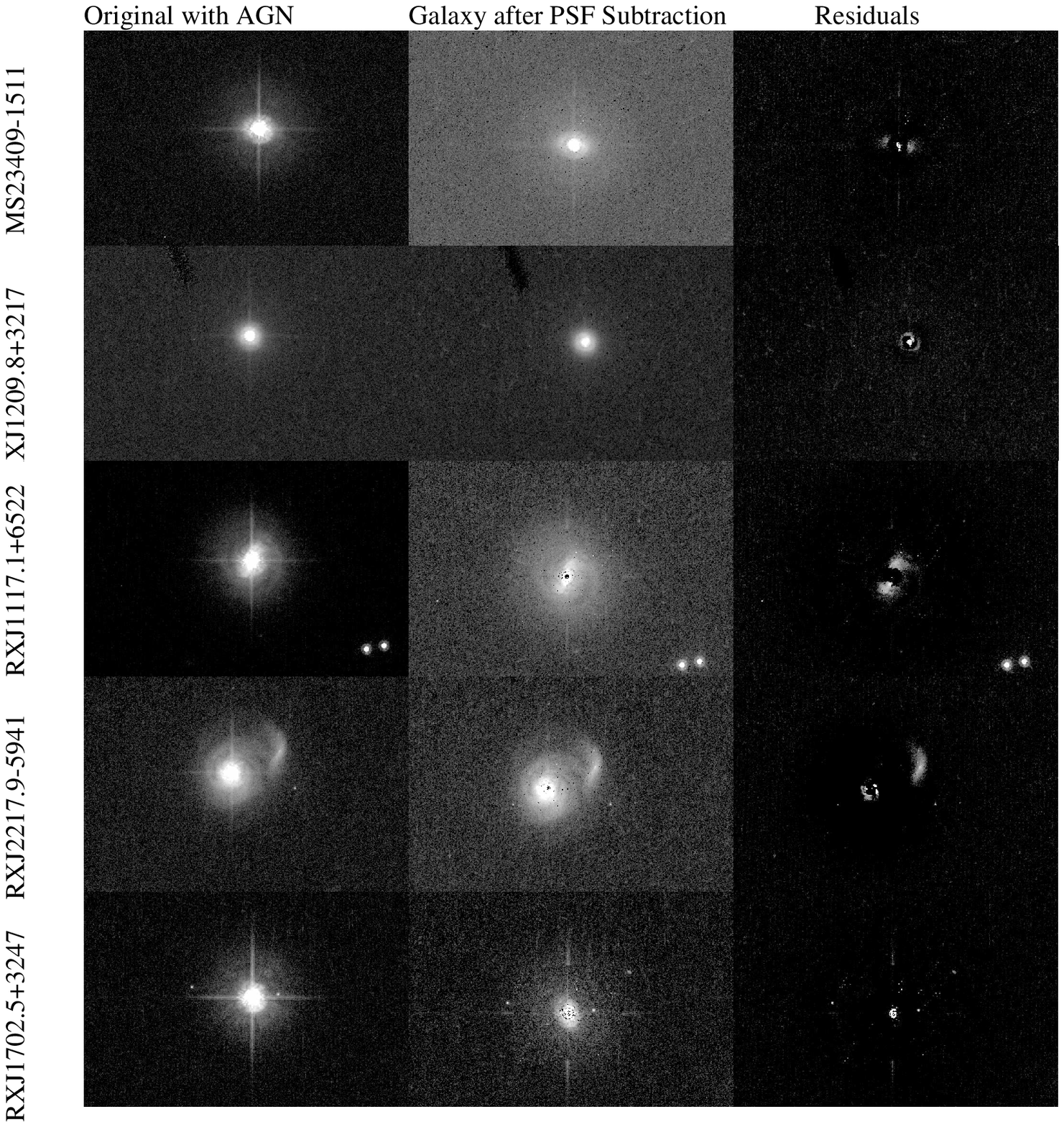,width=5in}
\caption{
Figure 1, continued. 
}
\end{figure}

\newpage

\begin{figure}
\psfig{file=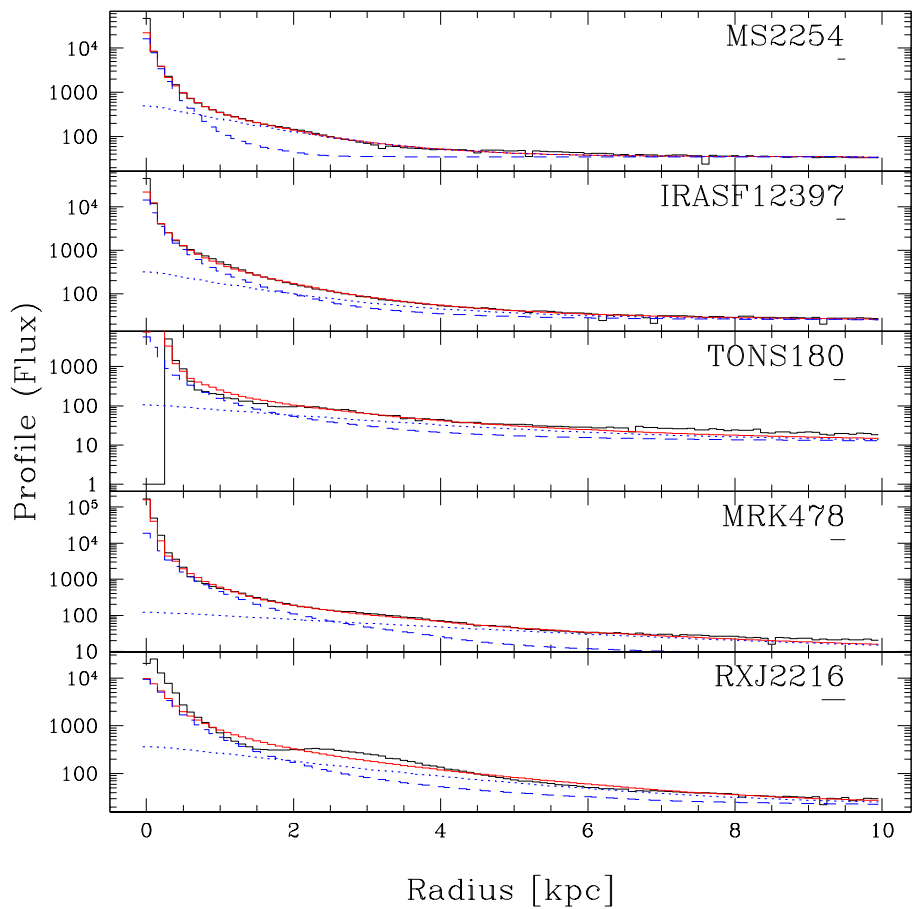,width=5in}
\caption{ Radial profiles of the sample galaxies. The dotted blue line
shows the disk component, while the dashed blue line shows the bulge
component and the solid red line is the sum of the two (the sky is
included in all). The black line is the data. The short horizontal bar
on the upper right corner (below the galaxy name) shows the size of the
PSF core (5 pixels). This shows that the galaxies are well sampled and
are well fit by the bulge$+$disk profile. }
\end{figure}

\begin{figure}
\psfig{file=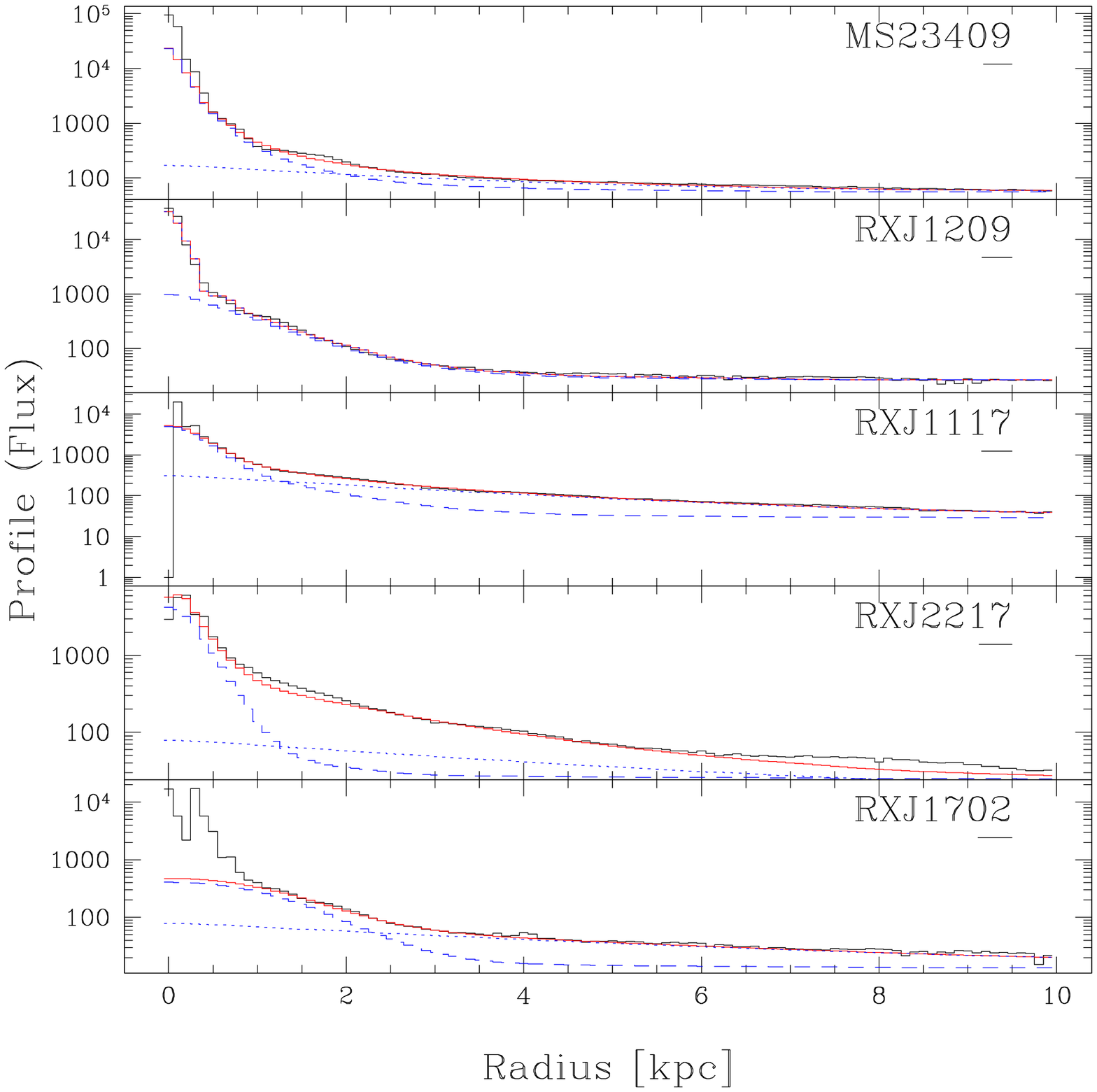,width=5in}
\caption{Figure 3, continued. }
\end{figure}


\begin{figure}
\psfig{file=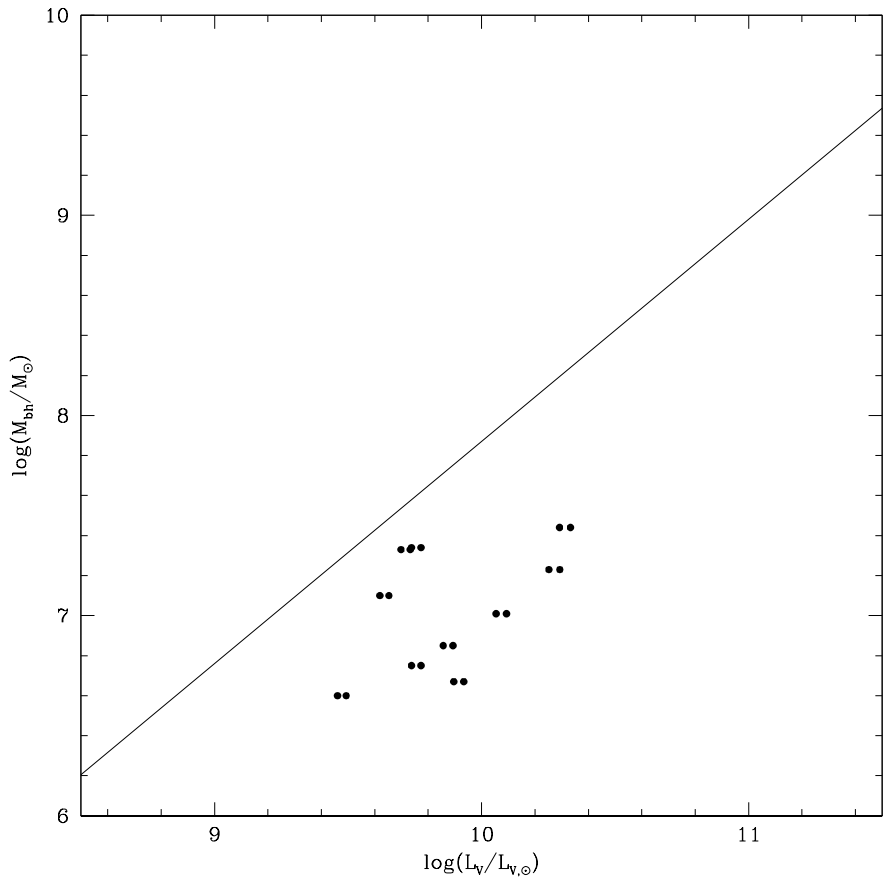,width=5in}
\caption{ The black hole mass vs. the host bulge luminosity for our
sample of NLS1s.  For each galaxy there are two points joined by a bar
corresponding to two different assumptions about the color corrections.
The line is the black hole mass-bulge luminosity relation from
G\"{u}ltekin et al. 2009.  It is clear that our sample galaxies do not
follow the G\"{u}ltekin et al, relation, but lie below that relation.
The measurement error on $\log L_V/L_{\odot}$ is smaller than the color
correction shown. The error on black hole masses estimates from single epoch
spectra is generally believed to be about 0.3 dex. }
\end{figure}

\newpage

\begin{figure}
\psfig{file=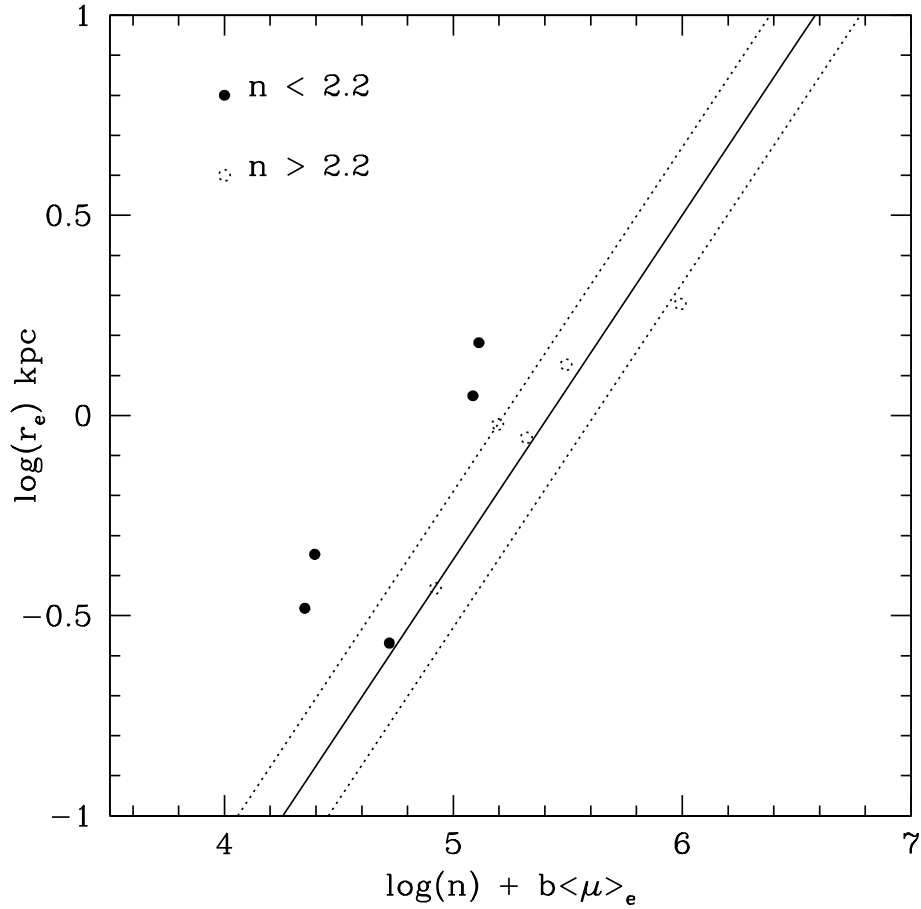,width=5in}
\caption{Photometric plane. The solid line is the best fit found for E
 \& S0 galaxies by Graham (2002). The points are our data. It appears
 that the pseudobulges do not lie on the photometric plane relation,
 just as they do not lie on the fundamental plane.}
\end{figure}

\end{document}